\documentclass[a4paper,10pt]{article}
\usepackage[english]{babel}
\usepackage{graphicx}
\usepackage[sorting=none]{biblatex}
\usepackage{color}
\usepackage{ccaption}
\usepackage{blindtext}
\usepackage{babel}
\usepackage{textcomp}
\usepackage[version=4]{mhchem}
\usepackage{gensymb}
\usepackage{geometry}
\usepackage{framed}
\usepackage{mathtools}
\usepackage{graphicx}
\usepackage{floatrow}
\usepackage{lineno}
\usepackage{csquotes}
\usepackage{silence}
\usepackage{amsmath,amsthm,amsopn,amstext,amscd,amsfonts,amssymb}
\usepackage[TS1, T1]{fontenc}
\DeclareUnicodeCharacter{17F}{{%
  \fontencoding{TS1}%
  \fontfamily{lmr}%
  \selectfont s%
}}

\begin{document}

\noindent \textbf{\huge Toward a general theory for the universality and scaling in critical thermal responses in biology}\\

\noindent José Ignacio Arroyo (1*)\\
Pablo A. Marquet (1,2,3,4)\\
Chris P. Kempes (1)\\
Geoffrey West (1)\\

\noindent (1) Santa Fe Institute, Santa Fe, New Mexico, United States, (2) Facultad de Ciencias Biol\'{o}gicas, Pontificia Universidad Cat\'{o}lica de Chile; CP 8331150, Santiago, Chile, (3) Centro de Cambio Global UC, Facultad de Ciencias Biol\'{o}gicas, Pontificia Universidad Cat\'{o}lica de Chile; CP 8331150, Santiago, Chile, (4) Instituto de Sistemas Complejos de Valpara\'{i}so (ISCV); Subida Artiller\'{i}a 470, Valpara\'{i}so, Chile.\\
\noindent *jiarroyo@uc.cl\\
\section*{INTRODUCTION}

Temperature is a major determinant of physical, chemical, and biological systems, and that includes socioeconomic systems. In biology, its effects are paramount since it modifies the reaction rates of enzymes, which regulate processes that manifest at all levels of biological organization \parencite{Dell2011}. Understanding its ecological and evolutionary role has attracted scientists for more than a century\parencite{margalef1951,larcher1973temperature, hochachka2002biochemical}. Empirical data show a high regularity in the temperature response across the entire range of biological phenomena from molecules to ecosystems and across taxa and environments \parencite{araujo2013heat, rezende2019thermal,schulte2011thermal,Dell2011,grimaud2017modeling,noll2020modeling,rezende2020predicting,yap2020predictive, kontopoulos2024no}. These temperature responses often have a concave asymmetric left-skewed shape (Figure 1, A), and less often a more or less symmetric or right-skewed shape (Figure 1, B-C), or a convex shape varying from left- to right-skewed (Figure 1, D-F). The properties of these responses have been shown to exhibit relationships among them \cite{rosso1993unexpected, huey1993evolution}, for example, between the maximum critical limit and optimum. Recently, many studies have studied the phenomena of the universality of thermal responses across levels of organization, biological quantities, and taxa \cite{luhring2017scaling, bozinovic2020thermal, kellermann2019comparing, rezende2019thermal, iverson2020thermal, dell2011systematic, febvre2024thermal, low2017predictions, rebaudo2018modelling, elias2014universality}, and the scaling of thermal properties \cite{peralta2021heat, lindmark2022optimum, knies2009hotter, angilletta2004temperature, leiva2019scaling}.
It is remarkable that despite the strong empirical foundation associated with the temperature dependence of biological processes, a simple first-principles theory has not yet been achieved (see Box 1). We developed a theory \parencite{arroyo2022general} showing that under appropriate normalizations and rescalings, temperature response curves show a remarkably regular behavior and follow a general, universal law. The impressive universality of temperature response curves remained hidden due to various curve-fitting models not well-grounded in first principles. In addition, this framework has the potential to explain the origin of different scaling relationships in thermal performance in biology, from molecules to ecosystems. Here we summarize the background, principles and assumptions, predictions, implications, and possible extensions of this theory.

\begin{framed}
\textbf{Box 1. The problem: Developing a simple, and general first-principles theory for temperature dependence, applicable from molecules to societies.} We lack a general first-principles theory to understand the temperature dependence of biological processes at different levels of biological organization, that it is simple enough to allow straightforward fitting to empirical data, is able to generate quantitative predictions,  and can be  integrated with other theories. For the case of temperature dependence, having a simple model describing the complete temperature response curve is fundamental to understanding the functioning of biodiversity from molecules to societies and predicting tipping points and the spread of diseases, for example, in response to global warming. The classical approach in biology is to use the Arrhenius equation, which only accounts for the exponential phase. Still, it does not allow for the prediction of tipping points, and extrapolating it can lead to erroneous assessments about the response of organisms to increases in temperature (see, for example, \cite{knies2010erroneous}). Biologists rarely have focused on understanding the origin and form of the other part of the response, the temperature decay phase of biological processes and rates, and most if not all models that have described the whole curved response have many parameters. Complicated models are often not easily tractable to integrate with other models or to derive simple analytical predictions. To fill this gap, we developed a simple theory based on first principles. \\
\end{framed}

\subsection*{Enzyme kinetics}
The basis of the temperature dependence of most if not all, biological rates, times, and quantities in general, is the temperature dependence of enzyme reaction rates constants ($k_1$, $k_{-1}$, and $k_{2}$).
The chemical equation that represents the reaction of a reactant S with an enzyme E to form a product P, passing by a transition state ES, is

\begin{equation}\label{mm}
\ce{E + S <-->[{$k_1$}][{$k_{-1}$}] ES} 
\ce{ <-->[{$k_{2}$}][{$k_{-2}$}] E + P}
\end{equation}

The Michaelis-Menten equation for the net chemical reaction rate, using  is and 'equilibrium approximation',

\begin{equation}\label{michaelis_menten_equilibrium}
\frac{dP}{dt}=v=\frac{k_2 e_0 S}{\left(\frac{k_{-1}}{k}\right)+S}
\end{equation}

Here $V_{max}=k_2 e_0$, $\frac{k_{-1}}{k_1}=K_d$ is the dissociation constant. The expression for $v$ can be simplified under certain assumptions. If $S>>K_d$ implies that,
\begin{equation}\label{assumption1}
v \approx k_2 e_0 \approx V_{max} 
\end{equation}

If $S<<K_d$ implies, 
\begin{equation}\label{assumption2}
v \approx \frac{k_2 e_0 S}{K_d}
\end{equation}

Below, we will show how temperature dependence relies on some of these simplifying assumptions, such as $S>>K_d$.

\subsection*{Models of temperature-dependence}
Most theories in biology regarding temperature dependence have been made using the Arrhenius equation: 

\begin{equation}
k=k_0 e^{-E/k_B T}
\label{basic-arrhenius}
\end{equation}

where $k$ is chemical reaction rate, $k_B$ is Boltzmann's constant, 
$T$ is absolute temperature, $E$ is an effective activation energy for the process of interest, and $k_0$ is an overall normalization constant characteristic of the process. This equation was originally an empirical formulation that was later motivated heuristically from chemical reaction theory (\parencite{laidler1983development}). For exampple, in the Metabolic Theory of Ecology (MTE), often also called Metabolic Scaling Theory or symply Scaling Theory, $k_0$ includes the body size, or mass ($M$) dependence, of the process and typically behaves as a power law, $M^{\beta}$, with the exponent $\beta$ being a simple multiple of $1/4$. For instance, for metabolic rate, $B$:

\begin{equation} 
    B=B_0 e^{-E/k_BT} M^{\beta}
    \label{arrhenius}
\end{equation}
where $B_0$ is independent of both $M$ and $T$, and the allometric exponent, $\beta$ $\approx$ $3/4$.
Consequently, a plot of $\log (B)$, or more generally, $\log (k)~vs.~1/T$ should yield a straight line, often referred to as an Arrhenius plot. More generally, in \ref{basic-arrhenius} (and \ref{arrhenius}), it has become a common practice in metabolic scaling theory (see for example \cite{brown2004toward,sibly2012metabolic}) to apply a logarithmic transformation and rearrange terms so as to obtain a "temperature corrected" metabolic scaling (or any other variable under study such as population density, productivity or diversity) ${\log (Be^{E/k_BT})=\beta\log(M)+\log(B_{0})}$ and a "mass corrected" metabolic rate scaling with temperature ${\log (BM^{-\beta})= -E(1/k_{B}T)+\log(B_{0})}$.

Although this phenomenological formula gives good agreement with data over the limited temperature range where rates increase with temperature, it fails badly when extended over the entire range by not predicting a critical temperature beyond which rates {\it decrease} with increasing temperature. On the other hand, the full expression for $k$ that we derive from first principles of chemical reaction rate theory,

\begin{equation}
k=k_0 T^{\gamma} e^{\Delta H/k_BT} 
\label{arrhenius2}
\end{equation}
where $\Delta H$ is the effective enthalpy of the underlying chemical reaction and the exponent $\gamma$ is given below (see Eq. \ref{eq:optimum}), automatically predicts such a critical (optimum) temperature, and leads to predictions in excellent agreement with data across the entire range of biological phenomena. Consequently, the correct expression that should  be used for the MTE in place of Eq.~(\ref{arrhenius}) is 
\begin{equation}
B=B_0 T^{\gamma} e^{\Delta H/k_BT} M^{\beta}
\label{arrhenius3}
\end{equation}
It should be emphasized that the "exact" equations (\ref{arrhenius2}) and (\ref{arrhenius3}) reduce to the Arrhenius relationships, Eqs. (\ref{basic-arrhenius}) and (\ref{arrhenius}), in certain limits, but, in general, should be treated as the replacement for the Arrhenius terms in any metabolic theory application. Below, we give details of the theory and highlight its successful confrontation with a plethora of data, but we begin with a short historical account.

The Arrhenius equation is a phenomenological function, but a similar mathematical form was later derived in different theories. Eq.~(\ref{basic-arrhenius}) was originally proposed as an empirical model by van’t Hoff (1884) \parencite{van1884etudes,laidler1984development}. Later in 1889, Arrhenius performed a data compilation of six studies that used different empirical equations to fit temperature dependence data of different chemical reactions (among them those of van’t Hoff) and showed that in all cases the data were well fit by Eq. \ref{basic-arrhenius}, expressed as $k=a e^{-b(1/cT)}$, where $a$ and $b$ are parameters, and $c$ is a constant. Later, following theoretical developments in statistical mechanics, thermodynamics, and collision theory, the equation was modified and re-interpreted; the b parameter was called the ``activation energy" (commonly denoted by $E$) and identified as the energy required to convert the inactive reactant into its active form, and $c$ was identified with the gas constant ($R$) or, equivalently, the  Boltzmann constant ($k_B$). 
Later, some derivations coincided mathematically with the Arrhenius equation but with a slightly different interpretation \parencite{laidler1983development}, though current use corresponds to the empirical formulation with the interpretation suggested by Arrhenius, as expressed in Eq.~(\ref{basic-arrhenius}).

More recently, in biology, and particularly in ecology, the Arrhenius equation has become a central component of the metabolic theory of ecology. In this theory, it has been applied to understand the temperature dependence of biological rates at different levels of biological organization, from molecules to ecosystems \parencite{Dell2011}. Although the Arrhenius equation has been instrumental in explaining the approximately universal temperature dependence of diverse biological rates \parencite{brown2004toward, Dell2011},  it cannot account for the complete pattern of temperature response of different biological traits, including metabolism and growth rate, among others \parencite{ratkowsky2005unifying, schulte2011thermal, Dell2011,price2012testing,knies2010erroneous, okie2012micro}. Experiments and observations have long established that the form of the temperature response has a (symmetric or asymmetric) concave upward or downward pattern relative to the canonical straight-line Arrhenius plot (e.g. \parencite{ratkowsky2005unifying}). 
Consequently, there are ranges of temperatures where the traditional Arrhenius expression, 
Eq.~(\ref {basic-arrhenius}), even gives the wrong sign for the observed changes in biological rates: namely, they {\it decrease} with increasing temperature rather than increase, as predicted by Eq. \ref{basic-arrhenius} for the {\it {entire}} temperature range. The Arrhenius model only accounts for the initial exponential phase of temperature responses. Consequently, biologists have commonly employed statistical distributions, such as the Laplace or Gaussian models, or higher degree algebraic polynomials (typically quadratic) \parencite{Dell2011}, to characterize the full performance curve. However, these too cannot accurately describe the observed temperature response since they are symmetric, in contrast to the data, which is often asymmetric. For ecology, the Arrhenius equation might be considered mechanistic in the sense that it gives a good phenomenological description of the temperature dependence of the relevant metabolic rates, i.e., of the dominant underlying enzymatic reactions. However, for physicochemists, it is generally viewed as an empirical equation that fits the data well \parencite{gutman2007empiricism}. This is often discussed, emphasizing the need to find a mechanistic model that also accounts for curvature \parencite{gutman2007empiricism,price2012testing,white2012information,white2012metabolic}.

There are, however, mechanistic first-principles models for understanding and calculating chemical reaction rates. These were originally formulated by Eyring and Polanyi, and by Evans, \parencite{eyring1935activated,evans1935some,arroyo2022general} and are usually referred to under the heading of  Transition State Theory (TST). Its extensions include several non-linear models \parencite{hinshelwood1946chemical,johnson1946growth,schoolfield1981non,peterson2004new,rabosky2018inverse,hobbs2013change,corkrey2014protein,schipper2014thermodynamic,delong2017combined,tang2023reanalysis,tang2024chemical}
which provide a fundamental basis for deriving the complete temperature dependence of biological rates and times. (For a review of these models, see \parencite{delong2017combined}.)\\
To summarise: Despite its importance, a comprehensive theory that unifies several key properties simultaneously has, up till now, not yet been developed.  Specifically, the key goals, which the framework we present here achieves, is a theory that: i) is based on first principles and fundamental physico-chemical mechanisms; ii) is simple in terms of its assumptions and mathematical form, yet efficient and parsimonious in that it explains a plethora of data and generates many predictions with few free parameters; iii) is general and applicable across multiple levels of biological organization and taxa, thereby manifesting a universal biophysical law.

\subsection*{Transition State Theory}
Here, we briefly introduce the Eyring-Evans-Polanyi (EEP) transition state theory (TST) \parencite{evans1935some}, which is the widely accepted theory of enzyme chemical kinetics. 
Below, we will show how TST can be simply extended to derive a simple general equation for temperature dependence for most biological rates and quantities. 
TST offers the possibility of developing a fundamental theory for the temperature dependence of biological processes that extends and generalizes the heuristic Arrhenius equation by grounding it in the first principles of thermodynamics, kinetic theory, and statistical physics \parencite{hanggi1990reaction,zhou2010rate}.

The framework of the TST 
conceives a chemical reaction as a flux of molecules with a distribution of energies and a partition function given by the Planck distribution, flowing through a potential energy surface (PES), which effectively simulates molecular interactions. The configuration of molecules flowing through this surface proceeds from i) a separate metabolite and enzyme to ii) an unstable metabolite-enzyme complex, which, iii) after crossing a critical energy threshold barrier, or transition state, then forms the final product (the transformed metabolite). A common misconception is that the ES complex is the transition state \cite{dalal2018:ch3}, but it is not. The chemical equation that includes a transition state includes, among others, the term $ES^{\ddagger}$, and additional rates $k_3$, $k_{-3}$, $k_4$, and $k_{-4}$ \cite{lienhard1973enzymatic,schramm1998enzymatic}. To use the simple equation for enzyme kinetics in equation \ref{mm}, it is necessary to make an assumption. If it is assumed that $k_3$, $k_4$ $>>$ $k_2$, the equation with a transition state \cite{lienhard1973enzymatic,schramm1998enzymatic} can be reduced to equation \ref{mm} \cite{johnson19921}. In what follows, we will assume this to derive a simple expression.

EEP thereby derived the following equation for the reaction rate:
\begin{equation}\label{eq: eyring1}
k=\frac{k_B}{h} T e^{-\Delta G/R T}
\end{equation}
where $h$ is Planck's constant, 
$\Delta G$ is the change in Gibbs free energy or free enthalpy, $R=N k_B$  is the universal gas constant, and $N$ is Avogadro’s number.  An overall coefficient of transmission is also originally part of Eq.~(\ref {eq: eyring1}) but is usually taken to be 1. Since the change in Gibbs free energy, or the energy available to do chemical work, can be expressed in terms of enthalpy ($\Delta H $) and the temperature-dependent change in entropy,  or dissipated energy ($\Delta S$), as $\Delta G=\Delta H-T\Delta S$,  Eq.~(\ref{eq: eyring1}) can then be written as \parencite{eyring1935a},

\begin{equation}\label{eqeyring2}
k=\frac{k_B}{h} T e^{\Delta S/R}  e^{-\Delta H/R T}
\end{equation}
Analogous to the Arrhenius expression, equations~(\ref{eq: eyring1}) and (\ref{eqeyring2}) describe an exponential response of the rate $k$ to the temperature provided, however, there is no temperature dependence of the thermodynamic parameters. Models have been developed for including this temperature dependence, but they typically invoke several additional assumptions and new parameters \parencite{daniel2001temperature,delong2017combined}. 
Moreover, most models for temperature response have been conceived for a single level of biological organization (primarily at the enzymatic/molecular level) \parencite{eyring1935activated,ritchie2018reaction} or for specific taxonomic groups; e.g. only for mesophilic ectotherms \parencite{Daniel2001}, endotherms \parencite{delong2017combined}, or thermophiles \parencite{prabhu2005heat}.\newline

\section*{THEORY}

\subsection*{Derivation of the theory}
Temperature changes the conformational entropy of proteins \parencite{wallin2006conformational}, which in turn determines the binding affinity of enzymes \parencite{Frederick2007,Tzeng2012} and affects the flexibility/rigidity and stability of the activated enzyme-substrate complex and hence the reaction rate [24]. The resulting temperature dependence of the change in entropy, $\Delta S$ (with enthalpy and heat capacity remaining constant), is the simplest mechanism for giving rise to curvature in an Arrhenius plot and naturally leads, via Eq.~(\ref{eqeyring2}), to power law deviations from the simple exponential form \parencite{sturtevant1977heat}. Following Prabhu and Sharp \parencite{prabhu2005heat}, the change of entropy for a given change in temperature can be expressed as,

\begin{equation}
T \frac{d\Delta S}{dT}=\Delta C
\end{equation}

, where $\Delta C$ is the heat capacity of proteins, assumed to be independent of temperature. Integrating over temperature gives 
\begin{equation}
\int^{T}_{T_0} d \Delta S= \int^{T}_{T_0} \frac{\Delta C}{T} dT
\end{equation}

\begin{equation}\label{eq:entropy}
\Delta S=\Delta S_0 + \Delta C\ln{(T/T_0)}
\end{equation}
where $\Delta S_0$ is the entropy when $T = T_0$, an arbitrary reference temperature, commonly taken to be 298.15 $K$ (25°C). 
Using this expression (\ref{eq:entropy}) for $\Delta S$ in Eq.~(\ref {eqeyring2}) and simplifying, we straightforwardly obtain:
\begin{equation}
k = {\frac {k_B}{h}} { {e^{\frac{\Delta S_0}{R}}}{T_0}^{\frac{-\Delta C}{R}}} 
\left( \frac {1}{T}\right)^{-({\frac{\Delta C}{R}} + 1)}
e^{\frac{-\Delta H} {RT}}
\label{EEP3}
\end{equation}
Here, $k$ is any of the rates in the chemical equation for enzyme kinetics. This has the form of a classic Arrhenius-like exponential term, modified by a power law, but with the “effective activation energy” is being replaced by the change in enthalpy. 

Under the assumption of $S>>K_d$, which implies that $v \approx k_2 e_0 \approx V_{max}$, we can simply use 
\begin{equation}
v \approx k_2 e_0 \approx k \approx {\frac {k_B}{h}} { {e^{\frac{\Delta S_0}{R}}}{T_0}^{\frac{-\Delta C}{R}}} 
\left( \frac {1}{T}\right)^{-({\frac{\Delta C}{R}} + 1)}
e^{\frac{-\Delta H} {RT}}
\end{equation}
Equation \ref{EEP3} describes cases of left-skewed and more or less symmetric temperature responses. However, as will be discussed in more detail in Box 3, the equation should be used in scale $\ln(Y)-1/T$.

\subsection*{Principles for applying the theory from molecular to macroscopic levels}

In order to apply the theory from the microscopic to the macroscopic scale, i.e., from quantum to classical scales, we need to introduce three important general principles, which can be summarized as follows: \\

\textbf{i)\textit {Averaging.}} Averaging emerges from summing. Metabolic rate, $B$, for example, is the sum of all reaction rates, $B=\sum k$. The sum is defined as $\sum k$, and an average as $\overline{k}=\sum k/n$, where $n$ is the number of reactions. Then $B= \sum k=\overline{k}n$. Metabolic rate is often simply defined as the oxygen consumed per minute or carbon dioxide produced per minute. In heterotrophic organisms, respiration is defined as $6CO_2 + C_6H_12O_6 \rightarrow 6CO_2+6H_2O$, and in autotrophic organisms, photosynthesis is defined as $6CO_2+6H_2O \rightarrow 6CO_2 + C_6H_12O_6$. In a heterotrophic cell, the reaction involving oxygen occurs at a final stage of the cell respiration during oxidative phosphorylation. In a single cell, several reactions occur simultaneously, then the metabolic rate is the sum os all those reactions. Moreover, for a multicellular organism, the sum of the sum must be taken.
There could be three alternative approaches to conceive the rate of an ensemble of enzymatic reactions: the rate-limiting reaction, a set of rate-limiting reactions, or the sum or average of the reactions.
Following Gillooly et al. \parencite{gillooly2001effects} and in the spirit of the MTE, we can extend our derivation from the microscopic up through multiple scales to multicellular organisms and ecosystems. Moreover, when calculating the averaging of a function $B(T)\approx B_0 \left( \frac {1}{T}\right)^{-({\frac{\Delta C}{R}} + 1)}
e^{\frac{-\Delta H} {RT}}$, the average results in an additional constant C, $B(T) \approx B_0 \left( \frac {1}{T}\right)^{-({\frac{\overline{\Delta C}}{R}} + 1)}
e^{\frac{-\overline{\Delta H}} {RT}}+C$. However, under certain assumptions, the constant can be neglected \parencite{savage2004improved}. This assumption is that if the variance in the parameters is small, then the average of a function describing a biological rate is approximately equal to the function of the average. Consequently, $B(T)$ can be approximated by an equation of the form of Eq.~(\ref{eq:master2}), but with the parameters being interpreted as corresponding averages. Metabolic rate, for example, can therefore be expressed as
$B(T) \approx B_0 \left( \frac {1}{T}\right)^{-({\frac{\overline{\Delta C}}{R}} + 1)}
e^{\frac{-\overline{\Delta H}} {RT}}$, where $B_0$ is a normalization constant.\\

\textbf{ii) \textit{Correspondence principle.}} 
This states that a more general theory must be reduced to a more specific theory in the appropriate limit.  
For instance, in physics, the classic example is that of quantum mechanics which, in the limit of Planck's constant 
$h \rightarrow 0$, must reduce to classical mechanics; i.e., Schr{\"o}dinger's equation must effectively reduce to Newton's laws when  $h \rightarrow 0$, which it does. Similarly, when velocities are small compared to the speed of light, special relativity reduces to classical mechanics. In our theory, this suggests that care has to be taken if we are to apply the same equation, namely Eq.~(\ref{EEP3}), at both macroscopic as well as microscopic levels since it explicitly contains  Planck's constant, $h$, in the denominator.

Its presence for microscopic enzymatic reactions appropriately reflects the essential role of quantum mechanics in molecular dynamics. On the other hand, for macroscopic processes, such as whole-body metabolic rate, the averaging and summing over macroscopic spatiotemporal scales, which are much larger than microscopic molecular scales, must lead to a classical description decoupled from the underlying quantum mechanics and, therefore, must be independent of $h$. This is analogous to the way that the motion of macroscopic objects, such as animals or planets, are calculated from Newton's laws and not from quantum mechanics and, therefore, do not explicitly involve $h$. Formally, the macroscopic classical limit is, in fact, realized when $h\rightarrow 0$. The situation here is resolved by recognizing that the partition function for the distribution of energies in the transition state of the reaction has not been explicitly included in Eq.~(\ref{eq:master1}). This is given by a Planck distribution, which includes an additional factor $(1- e^{-h\nu/k_BT})$, where $\nu$ is the vibrational frequency of the bond, as first pointed out by Herzfeld \parencite{herzfeld1919theorie}. For purely enzymatic reactions discussed above, this has no significant effect since $k_BT << h\nu$, and thus  $(1- e^{-h\nu/k_BT}) \rightarrow 1$,  resulting in Eq.~(\ref{eq: eyring1}). Multicellular organisms, however, correspond to the classical limit where $h\rightarrow 0$ so  $k_BT >> h\nu$ and 
$(1- e^{-h\nu/k_BT}) \rightarrow h\nu/k_BT$, thereby cancelling the $h$ in the denominator of Eq.~(\ref{EEP3}).
Consequently, the resulting temperature dependence of macroscopic processes, such as metabolic rates, become independent of $h$, as they must, but thereby lose a factor of $T$ relative to the microscopic result, Eq.~(\ref{EEP3}). So for example, for multicellular metabolic rate, $B$, this becomes: $B \approx  {B_0} \left(\frac{1}{T}\right)^{\frac{-\overline{\Delta C}}{R}} 
e^{\frac{-\overline{\Delta H}}{R T}}$, with the normalization constant, $ {B_0}$, no longer depending on $h$. Note that the above correction for the enzyme level can also be applied to Eyring's equations, Eq.~(\ref{eq: eyring1}) and (\ref{eqeyring2}), in which case they become mathematically identical to the Arrhenius relationship.
So, for example, following from the correspondence principle, both quantum and macroscopic rates can be combined into the single equation:\\
\begin{equation} \label{eq:master1}
Y \approx Y_0 (1/T)^{-(\Delta C/R)-\alpha} e^{-(\Delta H/R)(1/T)}
\end{equation}
where $Y_0$ is a reduced parameter, $\alpha=1$ at the enzymatic levels but $\alpha=0$ otherwise.\\

\textbf{iii)\textit {Extension beyond rates.}} The third assumption, and a corollary of the first, is that all rates whether transient, steady-state, or equilibrium, follow the same temperature response relationship. It is also important to note here that most biological quantities, which are not obviously rates themselves, are fundamentally associated with rates. This is true because biological quantities are either the integral of past rates, or are maintained by current rates \parencite{west2001general}. For example, diversity and abundance are in general terms functions of mutation rate, generation times, mortality rate, and energy requirements, and all these rates and times do vary with temperature \parencite{brown2004toward, allen2002global,allen2006kinetic,savage2004effects,bernhardt2018metabolic,gillooly2001effects}. An example in biology is the Verhulst model for population growth; where the rate $r$, equilibrium $K$ (and hence inflection point $K/2$) all depend on temperature \parencite {savage2004effects}.\\

\subsection*{General predictions of the theory}
Our model makes three sets of general predictions:\\

\textbf{i)\textit{General equation for a curved temperature dependence.}}\newline

Accordingly, our model can be applied from the micro to the macro, leading to a single master expression for the temperature dependence of any variable, $Y(T)$:
\begin{equation}\label {eq:master2}
Y(T) \approx  {Y_0} \left(\frac{1}{T}\right)^{\frac{-\overline{\Delta C}}{R} - \alpha} 
e^{\frac{-\overline{\Delta H}}{R T}}
\end{equation}
Here, $Y(T)$ represents either a rate, time, or transient/steady-state/equilibrium state \parencite{price2012testing}, and  $\alpha = 1$ for the molecular level and $0$ otherwise. It should be noted that the thermodynamic parameters may have additional implicit parameters (e.g. embodied in $Y_0$) that make the forms of Eqs.~(\ref{eq:master1}) and (\ref{eq:master2}) more complicated under certain conditions as, for instance, for reaction rates at the molecular level where $Y_0$ is determined by Eq.~(\ref{eq: eyring1}). \\

\textbf{ii) \textit{Analytic description of thermal performance and their scaling behavior}}\newline
One of the advantages of having a simple equation for the temperature response is that we can have a simple mathematical framework for different properties of the general equation (Eq. (\ref{eq:master1})) (zeros, inflection points, etc.), which facilitates the derivation of predictions. Below, we will briefly describe some of the properties of the function and the predictions they make regarding the curvature and limits.

A curve with negative curvature or concave is characterized by,

\begin{equation}
\Delta H>0, \Delta C<0 \rightarrow \frac{d^2 Y}{dT^2}>0 
\end{equation}

\begin{equation}
\lim_{T\rightarrow 0} Y(T)=0
\end{equation}

\begin{equation}
\lim_{T \to \infty} Y(T) = 0
\end{equation}

On the other hand, a curve with a positive curvature or convexity is characterized by,

\begin{equation}
\Delta H<0, \Delta C>0 \rightarrow \frac{d^2 Y}{dT^2}<0 
\end{equation}

\begin{equation}
\lim_{T\rightarrow 0} Y(T)=Y(T_0)
\end{equation}

\begin{equation}
\lim_{T \to \infty} Y(T) = \infty
\end{equation}

where $Y(T_0)$ is the value of Y when T=0.

Thermal performances of organisms do not reach the absolute zero or infinity and species often have well-established thermal minimum and maximum in the range -25 to 125 $\degree$ C, i.e. 248.15-398.15 K. Accordingly, to define these minimum and maximum temperatures in our equation we will define the minimum and maximum temperatures as the temperatures at which there is a very low Y. To do so we will first define the $Y_{max}$ as the maximum Y value, which we will show corresponds to $Y_{max}=Y(T_{opt})$ (where $T_{opt}$ is the inflection point), and will also define a very low or minimum $Y_{min})$ as the value of Y that is at or below the $5\%$ percent of the Y observed, $Y_{min} \leq \epsilon Y_{max}$, where $\epsilon$ is a small fraction; e.g. 0.05.
Accordingly, we can define the minimum and maximum temperatures by solving for T in the following equation,

\begin{equation}
Y_{m} \approx  {Y_0} \left(\frac{1}{T_m}\right)^{\frac{-\overline{\Delta C}}{R} - \alpha} 
e^{\frac{-\overline{\Delta H}}{R T_m}}
\end{equation}

where $T_m$ is the minimum or maximum temperature. Using the Lambert W function, we can obtain an analytic form for the minimum and maximum temperatures, $T_m$. The  Lambert W function is a multi-valued function such thatfor a function $y=xe^x$, $x=W(y)$. Using the W function to clear $T_m$,

\begin{equation}\label{Tm}
\frac{1}{T_{m}}=
\frac{\Delta C-\alpha R}{\Delta H}
W \left[\frac{\Delta H}{\Delta C-\alpha R}
\left(\frac{Y_{min}}{Y_0}\right)^{\left(\frac{1}{-(\Delta C-\alpha R)}\right)} 
\right]
\end{equation}

The Lambert W function has two branches $W_{-1}$ (if $-1/e \le x<0$) and $W_{0}$ (if $x \ge 0$) \cite{corless1996lambert}, then $T_m=T_{min}$ or $T_m=T_{max}$. In practice, it can be observed that the branch $W_{-1}$ gives $T_{min}$, $\frac{1}{T_{min}}=
\frac{\Delta C-\alpha R}{\Delta H}
W_{-1} \left[\frac{\Delta H}{\Delta C-\alpha R}
\left(\frac{Y_{min}}{Y_0}\right)^{\left(\frac{1}{-(\Delta C-\alpha R)}\right)} 
\right]$.
We can also obtain an expression for the inflection point by making $\frac{dY}{dT}=0$, $T_{opt}=\frac{-\Delta H}{(\Delta C-\alpha R)}$. The inverse will be,

\begin{equation}\label{eq:optimum}
\frac{1}{T_{opt}}=\frac{-(\Delta C-\alpha R)}{\Delta H}
\end{equation}

From combining equations \eqref{Tm},  \eqref{eq:optimum}, we  can predict linear relationships among (two of) the so-called cardinal temperatures, $T_{opt}$ and $T_{min}$.

It is possible to estimate other traits, such as the range, $T_{ran}=T_{max}-T_{min}$, and the left range, $T_{lran}=T_{opt}-T_{min}$. 

We can also derive the relationship between $Y(T_{opt})$ and $T_{opt}$, making in equation \eqref{eq:master2} $Y(T_{opt}) \propto (1/T_{opt})^{\frac{-\overline{\Delta C}}{R} - \alpha} 
e^{\frac{-\overline{\Delta H}}{R} (1/T_{opt})}$, and considering equation \eqref{eq:optimum}, $T_{opt}=
\frac{-\Delta H}{(\Delta C-\alpha R)}$, we obtain,

\begin{equation}\label{Yopt-Topt}
Y_{opt}=Y(T_{opt}) = Y_0 T_{opt}^{\Delta C/R}
\end{equation}

Given that $T_{opt}$ scales linearly with other traits $Y(T_{opt})$ should also scale with other tratis, such as $T_{min}$, for example.\\

iii) \textbf{\textit{Universal scaling and data collapse}}\newline

A classic method for exhibiting and testing the generality of an equation is to express it in terms of rescaled 
dimensionless variables, which predict that a plot of all of the data collapses onto a single "universal" curve [e.g. \parencite{west2001general}].
For simplicity we first define $a =\overline{\Delta C}/R + \alpha$, not to be confound the term $\alpha$ (alpha) (introduced in Eq. \ref{eq:master1}), and $b=\overline{\Delta H}/R$. To derive a dimensionless equation we divide Y(T) by the value of Y at the $T_{opt}$, $Y(T_{opt})$,

\begin{equation}
\frac{Y(T)}{Y(T^*)}=\frac{Y_0e^{\frac{-b}{T}}T^a}{Y_0e^{\frac{-b}{T_{opt}}}T_{opt}^a}
\end{equation}

We then define $\frac{Y(T)}{Y(T^*)}=Y^*$, and $\frac{T}{T_{opt}}=T^*$. Considering that $T_{opt}=-b/a$, then $b=-a/T_{opt}$, after replacing and rearanging,

\begin{equation}\label{eq:universal}
Y^{*}{^{-1/a}}=e^{-(T^{*-1}-1)}T^{*-1}=e^{-(1/T*-1)}1/T^{*}
\end{equation}

Our theory, therefore, predicts that when $Y^{*}{^{-1/a}}$ is plotted against $T^{*-1}=X^*$ all of the various quantities, regardless of the specific processes, collapse onto a single concave parameterless curve whose simple functional form is given by Eq.~(\ref{eq:universal}). Notice that this optimizes at $T^{*-1} = 1$ and encompasses in the same curve both the concave behaviors predicted in the original Arrhenius plot as a function of $T^{*-1}$. If we rise both sides of equation \ref{eq:universal} to the power $-1$, resulting in $Y^{*1/a}=e^{(T^{*-1}-1)}T^{*1}$ and plot $Y^{*1/a}$ vs. $T^{*1}$ the curve is convex.
We can also rewrite equation \ref{eq:universal} as,

\begin{equation}\label{universal2}
Y^{*}{^{-1/a}}=e^{-(X^*-1)}X^{*1}
\end{equation}

Notice that the new variable $X^*$ is the inverse of $T^*$,
$T^{*-1}=
\frac{1}{T*}=
X^*=
\left(\frac{T}{T_{opt}}\right)^{-1}=
\frac{1}{\left(\frac{T}{T_{opt}}\right)}=
\frac{T_{opt}}{T}=
\frac{\left(\frac{1}{T}\right)}{\left(\frac{1}{T_{opt}}\right)}$.
Eq. \ref{universal2} predicts a collapse when plotting $Y^{*-1/a}$ vs $X^*$. To obtain the curve for convex responses, we raise to the power $-1$ both sides of equation \ref{universal2}, obtaining  $Y^{*}{^{1/a}}=e^{(X^*-1)}X^{*-1}$. Then plotting $Y^{*}{^{1/a}}$ vs. $X^*$ results in a dimensionless convex curve with an optimum at -1. 
For the molecular level, $Y_0$ contains Planck's constant,$Y_0=Y_1/h$, but for the macroscopic level, it does not. This dimensionless equation predicts a generic concave and convex response with an inflection point at $(1,0)$.\\

We can take this one step further by noting that the function:

\begin{equation}
\tilde{Y}^{*}=
\frac{Y^{*-1/a}}{eT^{*-1}}=
e^{-T^{*-1}}
\label{eq:universal_exponential}
\end{equation}

, is predicted to be of a "pure" exponential Arrhenius form as a function of $T^*$. Thus, an even more dramatic manifestation of the universality and collapse of the data is to plot $\ln(\tilde{Y}^{*})$ vs. $1/T^*$ in a line with a negative trend. \\

Alternatively, the data collapse can be done with a power law, 
\begin{equation}
{\ddot Y^*} =\frac{Y^{*-1/a}}{e^{-T^{*-1}}e}=T^{*-1}
\end{equation}

Plotting $\ddot Y^*$ vs $T^{*-1}$, in log scale for example,$\ln(\ddot Y^*)$ vs $\ln(T^{*-1})$, results in a line with positive trend.



\begin{framed}

\textbf{Box 2. Scaling behavior in thermal performances.} Here, with "scaling behavior" we refer to some quantity $u$ that changes with a quantity $v$ according to $u=u_0 v^w$. Our theory predicts at least five classes of scaling behavior:

(i) In the general equation, in the limit $\Delta H \rightarrow 0$, it is predicted that 

\begin{equation}\label{powerlawdecay}
Y(T)\approx Y_0 T^{-(\Delta C+ \alpha R)}
\end{equation}

(ii) Regarding the most classical meaning of scaling in biology, the change of a biological rate or quantity with the size of a system (e.g. length, area, volume, or mass). One way to derive this is from our averaging approach, where the size is contained in the $Y_0$ term. So, if Y (e.g. species richness) and T (average temperature of a given area) remain constant, but the area of an ecological community changes, then it is expected that the thermodynamic parameter would change with different areas. There are empirical relationships in the literature, of relationships between thermal traits (e.g. minimum, range) and body size \cite{leiva2019scaling,fuller2016towards,peralta2021heat,riek2013allometry} or thermal conduction-related quantities and size \cite{mortola2013thermographic,kwak2016model}. As far as we know, there is no theory for the documented relationships between thermal traits and body size. We presume that our model could shed light on some of the principles that could originate those patterns. 
Besides theory development, more empirical studies are needed as most of these studies focus on a particular group (e.g. ectotherms) but not across distant-sized groups, e.g. bacteria and mammals.

(iii) By combining different equations that define the thermal traits, it is possible to predict linear scaling relationships among the thermal traits. For example, if we combine Eqs. (\ref{Tm}) and (\ref{eq:optimum}), it is predicted, under certain assumptions, that the optimum temperature should scale linearly with minimum temperature. Following this logic, it is possible to predict relationships among all thermal traits. Despite it has been reported that some traits for a given variable (e.g. population growth rate) are related among them \parencite{rosso1993unexpected}, it remains to be studied whether all traits are related and whether this is conserved or not for different variables. 

(iv) It can also be predicted relationships among $Y(T_{opt})$ and traits, e.g. $T_{opt}$. Despite the evidence for a relationship between $Y(T_{opt})$ and $1/T_{opt}$ and $1/T_{ran}$ in some taxa \parencite{huey1984jack}, it remains an open question whether this is true for all traits and in different taxa. \\
(v) There is a linear scaling relationship between the thermodynamic parameters $\Delta C-\alpha R$ and $\Delta H$, as predicted from \ref{eq:optimum}. If we plot $\Delta C-\alpha R$ vs. $\Delta H$ the slope should be $1/T_{opt}$. If we plot $\ln (\Delta C-\alpha R)$ vs. $\ln(\Delta H)$ the slope should be $1$.\\.

\end{framed}

\begin{figure}[ht!]
\centering
\centering
\includegraphics[width=0.45\linewidth]{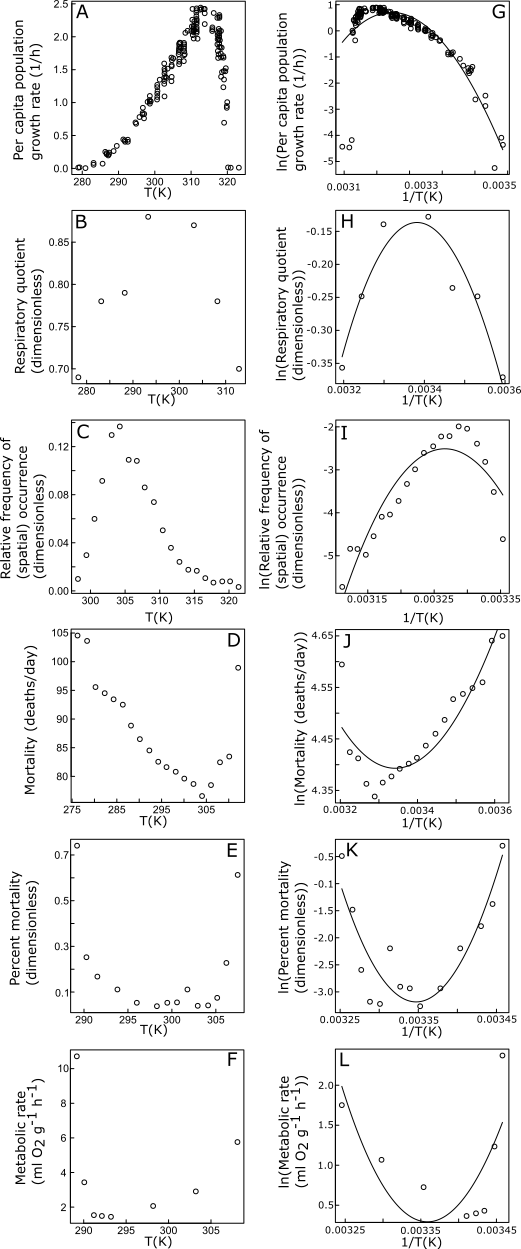}

\caption{\textbf{Examples of temperature response curves and their fit to equation (\ref{eq_log_scale}).} The plots in the left (A-F) are plotted in linear vs temperature (Kelvin), and the plots in the right (G-L) are plotted in logarithmic vs. $1/T$ (Kelvin). Plots A-C are concave, and D-F are convex. Plots A and D are left skewed, plots B and E are approximately symmetric, and plots C and F are right skewed. The data was compiled from \cite{ayala2022design,dingha2009effects,yang2020aggregation,lopez2021evolution,hargrove2020models, levin2015subtropical}. 
}
\label{fig2}
\end{figure}

\begin{center}
\fontsize{10pt}{12pt}\selectfont
\begin{tabular}{llllll}
\multicolumn{6}{l}{Table 1. Estimated parameters and goodness-of-fit statistics for fits in Figure 1}\\
\hline
Panel in Figure 1 & $\Delta C (JK^{-1})$  & $\Delta H (J)$ &$c(=\ln(Y_0))$&$r^2$&p-value\\
\hline
G & -8497.97 &2620958.12 &6890.18&0.68&<0.01\\
H & -1072.23&319527.81&869.7&0.9&<0.01\\
I & -25402.01&7779956.62&20558&0.84&<0.01\\
J & 756.37&-223839.15&-558.87&0.81&<0.01\\
K & 42283.77&-12630120.53&-34086.5&0.8&<0.01\\
L & 24190.41&-7199725.83&-19486&0.55&0.056\\
\hline
\multicolumn{2}{l}{}\\
\end{tabular}
\end{center}

\begin{framed}
\textbf{Box 3. Using the theory.} 
A simple way to use equation (\ref{eq:master2}) is fitting the reduced form in logarithmic scale,

\begin{equation} \label{eq_log_scale}
\ln(Y)=\ln(Y_0)-b(1/T)-a\ln(1/T)
\end{equation}

where $a=\Delta C/R+\alpha R$, and $b=\Delta H/R$. To do this we can use a non-linear regression model, such as the function "nlsLM" implemented in the R package minpack.lm. In practice, we fit the parameter $c=\ln(Y_0)$, which is simpler to fit than $Y_0$. An important point to mention regarding the fitting process is that some estimated values of $c$ can be very big or small, to calculate $Y_0$ from $c$, i.e., $e^c=Y_0$, and commonly, the estimation is zero or an infinite number. However, in the R environment, for example, there are specialized packages to calculate those values if needed, such as the "mpfr" function of the Rmpfr package.

Second, once the parameters have been estimated, we can compute directly the traits of the temperature response curve. The optimum is the ratio of $\Delta C-\alpha R$ and $\Delta H$ (Eq. \ref{eq:optimum}), however, the minimum or maximum would require to use the Lambert W function. Briefly, the W Lambert function is $W(y)=x \leftrightarrow y=xe^x$, has two solutions $W_{-1}$ and $W_{0}$. When fittign the function in scale $\ln-1/T$ the branch $W_{-1}$ calculates the $T_{min}$. For the decay of biological quantities with temperature, we can fit a power law of the form $Y=Y_0 T^{-a}$, or in ln scale,

\begin{equation} 
\ln(Y)=\ln(Y_0)-a\ln(T)
\end{equation}

For the relationships between thermal traits, we can also fit a power law of the general form $u=u_0 v^w$. Here $u$ and $v$ represent either a parameter or trait. The exponent $w$ is expected to be 1 or 0 depending on the relationship, and the prefactor $u_0$ also depends on the specific relationship.

After having estimated the parameters, the dimensionless form of our general equation can be used for data collapse to simply show that regarding different parameter values, all temperature responses have the same universal form. The dimensionless form can also be more conveniently used in log scale,

\begin{equation}\label{eq_log_scale_universal}
(-1/a)\ln(Y^{*})^{}=\ln(1/T^*)-[(1/T^*)-1]
\end{equation}

Eq. \ref{eq_log_scale_universal} predicts a concave curve with an inflection point at (1,0). If we multiply by $-1$, we obtain the equation for the convex curve, 

\begin{equation}\label{eq_log_scale_universal_convex}
(1/a)\ln(Y^{*})^{}=-\ln(1/T^*)+[(1/T^*)-1]
\end{equation}

\end{framed}

\subsection*{Empirical support for the theory}

As is traditional in thermal biology and metabolic ecology, we compare predictions from our theory with a semi-log plot of the data, i.e., $\ln(Y)$ vs. $(1/T)$, inspired by the original Arrhenius equation, Eq.~(\ref{arrhenius}). In this scale, our model predicts a curved temperature response, according to Eq. \ref{eq_log_scale}. (Box 3) Also importantly, in this scale, our model is simple to implement in a non-linear regression model and estimates not just the parameters but the different thermal traits (see Box 3). Our theory provides an excellent fit to a wide variety of temperature response variables across different taxa, including both concave and convex, and left-skewed, symmetric, and right-skewed patterns. In Figure 1 (G-L), we show some representative examples of fits to concave patterns with long tails at low and high temperatures, as well as convex patterns, such as the effect of environmental temperature on endotherm metabolism (in torpor) and biological times, also with tails at both ends. The fit to the data shows that the model is good at fitting curves, whether concave or convex and more or less symmetrical, but it is not good at fitting the extreme decay of asymmetrical patterns. The estimated parameters for the fits shown in Figure 1 are in Table 1.
Our prediction (\ref{eq_log_scale_universal}, \ref{eq_log_scale_universal_convex}) of the universal curve is very well supported by data, as illustrated in Fig. 2, where the collapse of all of the data from this study for both convex and concave patterns, regardless of organizational level, temperature range, or taxa, is shown. This result strongly supports the idea that our theory captures all of the meaningful dimensions of thermodynamic and temperature variation for diverse biological properties, which, when appropriately rescaled, can ultimately be viewed as a single simple exponential relationship, Eq.~(\ref{eq:universal_exponential}).
We also found some examples in the literature of the hypothesized scaling relationships in thermal curves (Box 2; Fig. 3). These include an example of the power-law decay of a biological quantity (species richness in hot springs, environments characterized by high temperatures) and temperature, the relationships among thermal traits (e.g., maximum and optimum), maximum population growth rate and optimum temperature, and optimum temperature and body mass.\\

\begin{figure}[ht!]
\centering
\centering
\includegraphics[width=0.68 \linewidth]{fig2definitive.png}

\caption{\textbf{Universal patterns of temperature response predicted by Eq. \ref{eq_log_scale_universal} (A) and \ref{eq_log_scale_universal} (B).}}
\label{fig3}
\end{figure}
\newpage

Figure 2 (Continuation). The upper panel (A) shows the concave and the bottom panel (B) the convex non-linear patterns predicted when $\ln (Y^*)^{-1/a
}$ is plotted vs.~$1/T^*=X^*$, [Eq.~\ref{eq_log_scale_universal}]. In theory, if all data is rescaled using the estimated parameters according to the same equation and the data collapses into a single pattern, it means that all respond to the same general law. As expected, all curves, regardless of variable, environment, and taxa, collapse onto a single curve when plotted in either of these ways. Notice that the pattern described by the universal curve is asymmetric. This is evident when observing that (i) for the concave universal curve, at the left side, $1/T^*<0.8$ when $\ln(y^*)^{-1/a}=-0.02$, but at the right $1/T^*>1.2$. If the curve were symmetric should be <1.2. (ii) For the convex universal curve, at the left side, $1/T^*<0.8$ when $\ln(y^*)^{-1/a}=0.02$, but at the right $1/T^*>1.2$. If the curve were symmetric should be <1.2.
Also, notice that the data is plotted in (dimensionless) inverse temperature, then the right tail in inverse temperature corresponds to the left tail on the temperature scale. Only those fits with a p-value < 0.05 were included in the data collapse. The data correspond to a database compiled in \cite{arroyo2022general}.

\begin{figure}[ht!]
\centering
\centering
\includegraphics[width=0.7\linewidth]{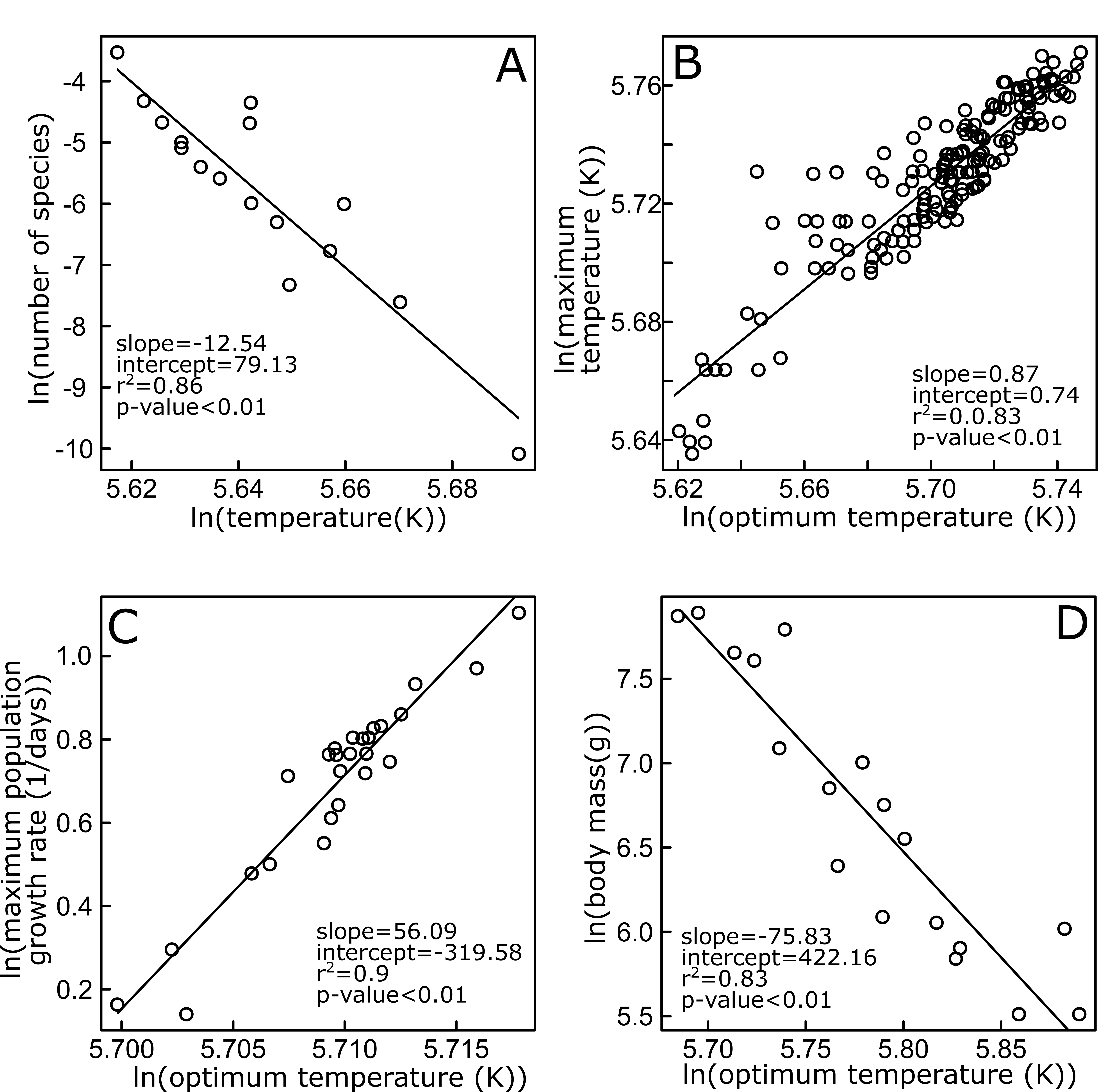}
\caption{\textbf{Examples of scaling behavior in thermal performance} (A) Relationship between species richness and temperature in hot springs \cite{ruhl2022microbial}, (B) maximum and optimum temperature in phytoplankton, insects, and lizards \cite{pinsky2019greater}. (C) population growth rate and optimum temperature in \textit{Drosophila} \cite{alruiz2023temperature} and \textit{Paramecium}\cite{krenek2012coping}.(D) optimum temperature and body mass in insects \cite{angilletta2004temperature}. Abbreviations: r-sq: r-squared, p-val: p-value.}
\label{fig4}
\end{figure}

\section*{IMPLICATIONS AND EXTENSIONS}

Our equation for temperature response, derived from underlying principles of chemical reaction rate theory, supersedes, and should therefore be used instead of the Arrhenius equation, Eq.~(\ref{arrhenius}), which is unable to account for the curvature in classic semi-log Arrhenius plots. As an example, and as has already been emphasized, it extends the validity of the master equation of the MTE \parencite{brown2004toward} by generalizing it to the expression:

\begin{equation}
Y(M, T) \approx  {Y_0}M^{\beta} \left(\frac{1}{T}\right)^{\frac{-\overline{\Delta C}}{R} - \alpha} 
e^{\frac{-\overline{\Delta H}}{R T}}
\end{equation}

It is relevant to remark that for scaling to body size in biology, this model could be used for normalization and provide probably a better option to decrease unexplained variance, using the normalization,
$Y(M, T)\left(\frac{1}{T}\right)^{\frac{\overline{\Delta C}}{R} - \alpha} 
e^{\frac{\overline{\Delta H}}{R T}} \approx  {Y_0}M^{\beta}$.
An important application of this theory is that it gives a principled, yet
simple mathematical tool for making predictions for the effects of climate change, and in particular, for predicting inflection points and whether quantities either increase or decrease with increasing temperature. For instance, it could be used to estimate the relationship between temperature change and various socio-economic quantities related to sustainability, such as electricity consumption \parencite{Yi-Ling2014influences}. 
More complex assumptions could lead to the inclusion of additional parameters that could help further explain the variance in temperature response curves. Among these possibilities are, for example, to consider alternative assumptions in the Michaelis-Menten equation.
Other possibilities can include assuming a non-linear temperature dependence of other thermodynamic parameters, such as heat capacity (see, for example, \parencite{yeh2023implications}).
Also, further extensions of this theory might include incorporating additional variables, such as pH, salinity, and oxygen availability; however, this may be important, especially in aquatic environments. Consequently, these effects have become a priority for increased theoretical and empirical research in the context of current global changes \parencite{portner2001climate,hochachka2002biochemical,somero2012physiology, somero2017biochemical}.\\
It should be borne in mind that our theory is based on equilibrium thermodynamics, and as such is not designed to deal with fluctuations out of equilibrium, nor with temporal changes in state variables, for example due to acute exposure or different exposure times to a given temperature \parencite{kingsolver2016beyond,jorgensen2019assess,rezende2020predicting}. However, our framework can in principle deal with slow changes relative to fast microscopic changes, assuming quasi-equilibrium (or quasi-static) conditions, or a slowly changing temporal sequence of equilibrium states \parencite{schroeder2012introduction}. Future connections with non-equilibrium thermodynamics may prove valuable in this regard \parencite{demirel2018nonequilibrium}.\\
As already pointed out, due to its mathematical simplicity, this theory is easily extendable to explain other patterns such as relationships among different attributes of the thermal response (e.g., \parencite{dixon2009relationship})
since these traits depend on the same parameters, such as the maximum value of the dependent biological quantity and its range. This could potentially shed light on several biological hypotheses such as "hotter is better" \parencite{angilletta2010thermodynamic,huey1989evolution} (i.e., a positive relationship between the maximum value of a dependent quantity and its optimum temperature), or "jack-of-all-temperatures, master of none" \parencite{huey1984jack,gilchrist1995specialists,huey1989evolution} (i.e., a trade-off between the maximum value of a dependent quantity and the breadth of performance).\\ 
Finally, it is worth pointing out that the theory can be extended to determine the temperature dependence of other relevant thermodynamic variables such as mass, volume, density, and pressure. 
For example, in the context of global warming, an important challenge is to be able to predict the sensitivity of thermodynamic parameters (such as enthalpy and heat capacity) of organisms of different sizes to temperature increases.\\

\section*{CONCLUSIONS}

Here, we have developed a theory for critical thermal phenomena in biology that makes three general predictions: i) a left-skewed or (almost) symmetric curved temperature response of biological quantities, that can be fitted to data in $\ln(Y)-1/T$ scale, ii) a simple framework for the thermal properties of organisms, leading to predictions such as scaling behavior of thermal properties, e.g. between maximum and minimum temperatures, and iii) a universal scaling and data collapse that describes the temperature dependence as a universal law across all levels of biological organization, taxa, and the whole range of temperature within which life can operate (approximately $-25$ to $125$ $\degree C$). Beyond these basic predictions, our frameworks provide a basis to derive the origin of different scaling behaviors observed in thermal responses, including the relationships between thermal traits and performance and body size, for example.
Importantly, our framework can be used for predicting scenarios of global warming, disease spread, and industrial applications. It provides a simple general equation that can be readily integrated into different theories in ecology and evolutionary biology.\\

\subsection*{Acknowledgements} We dedicate this chapter to the memory of our friend Francisco Bozinovic.  J.I.A. was supported by a Beca de Doctorado Nacional Agencia Nacional de Investigacion y Desarrollo (ANID) Grant 21130515. P.A.M. was supported by Grants ANID-Fondo de Desarrollo Cientifico y Tecnologico (FONDECYT) 1200925 and Exploracion 13220168 and by Centro de Modelamiento Matematico (CMM), Grant FB210005, and Inria Challenge project OceanIA (desc. num 14500). J.I.A. and
G.B.W. was supported by NSF Grant 1838420, J.I.A. and C.P.K. were supported by NSF Grant 1840301, and G.B.W. and C.P.K. were supported by the Charities Aid Foundation of Canada for the grant entitled “Toward Universal Theories of Ecological Scaling.” \\
\subsection*{Code availability.}
Data and codes are available at \sloppy 
\url{https://github.com/jose-ignacio-arroyo/tem.mod}
\printbibliography

\end{document}